\documentclass[12pt]{article}
\usepackage{hyperref}   
\hypersetup{
    colorlinks=true,
    linkcolor=blue,
    filecolor=magenta,      
    urlcolor=blue,
}
\usepackage{url}          
\usepackage{authblk}

\begin{document}

\title{Room Temperature, Quantum-Limited THz Heterodyne Detection?
Not Yet}
\author[1]{J. Zmuidzinas}
\author[2]{B. Karasik} 
\author[3]{A. R. Kerr}
\author[3]{M. Pospieszalski}
\affil[1]{California Institute of Technology 301-17, 
1200 E. California Blvd. , Pasadena CA 91125. jonas@caltech.edu}
\affil[2]{Jet Propulsion Laboratory 168-314, California Institute of Technology, 
4800 Oak Grove Drive, Pasadena CA 91109 }
\affil[3]{National Radio Astronomy Observatory, 
1180 Boxwood Estate Road, Charlottesville VA 22903}

\maketitle

\noindent Arising from N. Wang et al., Nature Astronomy (2019)\\
\url{https://doi.org/10.1038/s41550-019-0828-6} 

In their article, Wang et al. \cite{Wang2019} report a new scheme for THz heterodyne detection using a laser-driven LTG-GaAs 
photomixer \cite{brown1995photomixing,berry2013significant} 
and make the impressive claim of achieving near quantum-limited sensitivity
at room temperature.
Unfortunately, their experimental methodology is incorrect, 
and furthermore the paper provides no information on the
mixer conversion loss, an important quantity that could readily
have been measured and reported as a consistency check.
The paper thus offers no reliable experimental evidence that
substantiates the claimed sensitivities. To the contrary, 
the very high value reported for their photomixer impedance
strongly suggests that the conversion loss is 
quite poor and that the actual sensitivity
is far worse than claimed.

THz heterodyne detection has been
used in astronomy for over three decades \cite{boreiko1988heterodyne} and
is primarily of interest for high-resolution spectroscopy \cite{zmuidzinas2004superconducting}.
An illustrative application is the measurement of rotational
transitions of isotopologues of water vapor to estimate the D/H
ratio in comets -- indeed, recent data from the
SOFIA airborne observatory have reactivated the debate on whether the water in the Earth's oceans was delivered by comets \cite{lis2019terrestrial}.
To date, the best detection sensitivities have been achieved using
cryocooled superconducting devices \cite{zmuidzinas2004superconducting} such as the superconductor-insulator-superconductor (SIS) receivers
that enable ALMA \cite{wootten2009atacama} or the hot-electron bolometers (HEB) used in the HIFI instrument
on the  Herschel Space Observatory \cite{pilbratt2010herschel}.
However, other THz space instruments \cite{siegel2007thz}   
have generally not used cryocooled detectors 
due to power, mass, and cost constraints,
thereby incurring a sensitivity penalty of $10\times$ or more.
If the sensitivities reported by Wang \textit{et al.}
were correct, this severe penalty for ambient-temperature
operation would be erased and numerous applications would be enabled.

The sensitivity of a heterodyne receiver
is usually reported as a double sideband (DSB) receiver noise 
temperature
$T_\mathrm{rec}$.
By definition, $T_\mathrm{rec}$ is the Rayleigh-Jeans temperature
of a perfect blackbody source that,
when illuminating the receiver's THz-frequency input,
would contribute the same noise power
at the output of the final IF amplifier (typically in the GHz range)
as contributed by all of the different components of the THz receiver.
The sensitivity of a heterodyne receiver is subject to the quantum limit \cite{caves1982quantum} 
arising from the Heisenberg uncertainty principle, corresponding to
$T_\mathrm{rec} \ge h\nu/2k$ for the DSB noise
temperature.
As shown in their Figure~3c,
the sensitivity claimed by Wang \textit{et al.} 
for their room-temperature receiver
lies impressively close to this limit,
matching or exceeding the performance of
cryocooled SIS or HEB receivers.

Unfortunately, the experimental procedure
used by Wang \textit{et al.}
to measure sensitivity is fatally flawed.
The standard technique for determining the noise temperature is 
the ``$Y$-factor'' method where the IF noise power
$P_\mathrm{IF}$ is measured  using two blackbodies with
different Rayleigh-Jeans temperatures, $T_\mathrm{hot}$ 
and $T_\mathrm{cold}$.
The $Y$-factor is the ratio of these two measurements,
\[
Y = \frac{P_\mathrm{IF,hot}}{P_\mathrm{IF,cold}}
\]
from which the receiver sensitivity is calculated 
according to
\[
T_\mathrm{rec} = \frac{T_\mathrm{hot} - T_\mathrm{cold}}{Y-1} - T_\mathrm{cold}\ .
\]
In their Methods section, Wang \textit{et al.} 
claim to measure sensitivities  ``using the standard $Y$-factor method'', but this is not the case.
In that section, we learn that: 
\begin{quote}
``...the output IF signal at $\sim 1$~GHz 
is detected by a power meter (Mini-Circuits ZX47-60LN) 
using a lock-in amplifier with the 100~kHz modulation
reference frequency...''.
\end{quote}
The output of the ZX47-60LN power detector
\cite{MiniCircuits}
is a voltage that responds logarithmically to the
IF power $P_\mathrm{IF}$; one could presumably use the measured DC voltage
along with a calibration curve to obtain $P_\mathrm{IF}$,
as needed for calculation of the $Y$-factor.
However, a lock-in amplifier does not measure the DC
voltage; rather, it measures \textit{changes} in the voltage
that occur at a 100~kHz rate in response to a modulation.
Their Methods section states
\begin{quote}
``...the optical pump beam from the dual DFB laser system is modulated using an acousto-optic modulator ... at a 100 kHz rate''
\end{quote}
meaning that the laser source that drives the photomixer is being
turned on and off at the 100~kHz lock-in frequency.
Thus, the lock-in reports a quantity that is related to the
change in the IF power that occurs in response to turning
the laser on and off,
$\Delta P_\mathrm{IF} = P_\mathrm{IF,on} - P_\mathrm{IF,off}$.
Given their statement regarding use of the $Y$-factor method,
Wang \textit{et al.} presumably take the ratio of 
such lock-in readings for hot and cold loads.
Although not explicitly stated in their paper, 
they apparently make the assumption that the response of 
their logarithmic ZX47-60LN detector is linear for small perturbations, in which case
taking the ratio yields
\[
Y' = \frac{P_\mathrm{IF,hot, on} - P_\mathrm{IF,hot,off}}
{P_\mathrm{IF,cold,on} - P_\mathrm{IF,cold,off}}\ .
\]
This quantity is very different than the standard $Y$ factor.
In particular, $Y'$ remains invariant if a constant is added to
all power levels, e.g., as a result of a uniform increase of
the IF amplifier noise contribution, whereas $Y$ is definitely not invariant.
Furthermore, the paper makes no mention of measurements
of $P_\mathrm{IF,hot,off}$ or $P_\mathrm{IF,cold,off}$,
and because this information is missing, it is not possible
to convert the reported values of $Y'$ into corrected
values for $Y$. 
Noise temperatures calculated using $Y'$ are meaningless,
and can be radically different from the
true noise temperatures calculated using $Y$.
Thus, one cannot place any confidence in the 
sensitivities reported by Wang \textit{et al.}
in their Figure~3c.

What would a properly executed $Y$-factor measurement reveal
about the performance of their device?
Judging from the large photomixer impedance reported
by Wang \textit{et al.},
the sensitivity is likely $\sim 100\times$ worse than claimed.
Our argument involves the mixer conversion loss $L$,
which has a direct impact on receiver 
sensitivity $T_\mathrm{rec}$ according to
\[
T_\mathrm{rec} = T_\mathrm{mixer} + L T_\mathrm{IF\, system}\ .
\]
The mixer noise temperature $T_\mathrm{mixer}$ cannot be negative,
and the noise temperature of the first-stage low-noise amplifier
$T_\mathrm{LNA}$ sets a lower bound for the noise 
of the IF system, $T_\mathrm{IF\, system}$.
Thus, $T_\mathrm{rec} \ge L T_\mathrm{LNA}$ must hold.
Unfortunately, Wang \textit{et al.} do not report any values
for the conversion loss $L$, apparently missing several opportunities
for measuring $L$ while collecting data for their figures 2 and 3.
Wang \textit{et al.} also do not report measurements
for $T_\mathrm{IF\, system}$ or $T_\mathrm{LNA}$,
although according to the manufacturer's 
data sheet for their Mini Circuits ZRL-1150
first-stage amplifier \cite{MiniCircuits}, we may take
$T_\mathrm{LNA} \ge 70$~K as a reasonable value including cable losses.
Thus, the $\sim 150$~K noise temperatures reported by 
Wang \textit{et al.}
at frequencies up to 1~THz require a conversion loss no higher than
$L \approx 2$, or 3~dB. 
This is an extremely low value, comparable to the conversion loss of
a well-optimized SIS mixer;
if their 300~K mixer contributes any noise at all, the conversion
loss would need to be even lower.

The actual conversion loss is likely
$\sim 20$~dB worse,
as indicated by the photomixer impedance
reported by Wang \textit{et al.} on page 8 of the supplemental information:
\begin{quote}
``...$R_P$ is the average electrical resistance of the photomixer at a 30 mW pump power. 
The estimated $R_P$ value from the numerical simulations is in agreement with the
experimentally measured photomixer resistance of 
$25 \, \mathrm{k}\Omega$.''
\end{quote}
This large impedance is in line with data on other LTG-GaAs photomixers
including those previously studied by this group \cite{berry2013significant}.
Thus, there is a severe impedance mismatch between the
$25 \, \mathrm{k}\Omega$ photomixer and the
$50 \, \Omega$ IF amplifier,
corresponding to a coupling loss of 21~dB.
For the overall conversion loss to be no worse
than 3~dB, as required by the reported sensivity,
the photomixer would need to have an internal conversion gain
of at least 18~dB! 
Internal conversion gain in this device seems implausible, 
especially such a large value, 
given the lack of any measurements of conversion loss
and the absence of a clear physical mechanism for gain.

Indeed, the theory of operation presented by  Wang \textit{et al.}
makes no mention of conversion gain.
According to equation~(6) of their supplemental
information, they assume a linear, local relationship between
the current density $\vec J$ and THz electric field $\vec E$
at frequency $f_\mathrm{THz}$.
For $f_\mathrm{THz} < 1 / 2 \pi \tau \approx 500$~GHz
where $\tau = 0.3\,$ps is the stated carrier lifetime,
the response is essentially instantaneous and their equation~(6)
is equivalent to Ohm's law 
$\vec J(\vec r, t) = \sigma(\vec r, t) \vec E(\vec r, t)$.
Here $\sigma(\vec r, t)$ is the (real) conductivity
that varies with time $t$ and position $\vec r$
due to the photogeneration of carriers by the two
lasers with beat frequency $f_\mathrm{beat}$.
The equation $\vec J = \sigma \vec E$ simply describes 
current flow in an ordinary resistor, 
and leads to the circuit version of Ohm's law 
$I = V/R$, or $I(t) = V(t)/ R(t)$
when the conductivity is time dependent. 
Here $I(t)$ and $V(t)$ are
the current and voltage across the terminals of the THz
spiral antenna, and the $R(t)$ is the time-dependent
photomixer resistance as seen from the antenna terminals.
Thus, the theory offered by Wang \textit{et al.}
places the device into a well-known class 
of ``resistive mixers'' that includes 
diode mixers \cite{torrey1948rad} and 
FET mixers \cite{peng1997},
which are not capable of conversion gain
and in fact are subject to a theoretical minimum
conversion loss of 3~dB
\cite{kelly1977fundamental,tucker1985quantum}.
Their theory is therefore incompatible with their claimed sensitivity, which requires a large internal conversion gain
to overcome the severe IF impedance mismatch
as discussed above.
While the resistive mixer argument strictly holds only
for $f_\mathrm{THz} < 500$~GHz,
one expects photomixer performance to 
deteriorate at higher frequencies \cite{brown1995photomixing},
as Wang \textit{et al.} themselves state
when discussing the utility of the short lifetime $\tau = 0.3$~ps to
``recombine the slow photocarriers that degrade the terahertz-to-RF
conversion efficiency''.

We close by offering recommendations
the authors could adopt to address our
concerns and to build confidence in their results:
\begin{itemize}
\item Eliminate the lock-in and 
use the standard $Y$-factor method to determine $T_\mathrm{rec}$.
It may be helpful to add IF gain to boost the detector output
and to use a square-law IF detector rather than a logarithmic detector.
\item Measure and report the mixer conversion loss,
an urgently needed consistency check. This could be measured
during the hot/cold load procedure if the IF system is calibrated.
\item Measure the noise temperature of the IF system, and estimate
the IF contribution to the receiver noise using the measured conversion loss.
\item Experimentally demonstrate
that the receiver response is truly heterodyne,
e.g., by using infrared-blocking filters, THz passband filters,
and ideally gas-cell measurements of molecular absorption lines.
%
%
%
\end{itemize}

\noindent
The research by BK was performed at the Jet Propulsion Laboratory, 
California Institute of Technology, under a contract with the National Aeronautics and Space Administration.
The National Radio Astronomy Observatory is a facility of the National Science Foundation operated under cooperative agreement by Associated Universities, Inc.

\bibliography{ZKKP.bib}

\end{document}